\documentclass[preprint,12pt]{elsarticle}

\usepackage{amssymb}
\usepackage{amsmath}
\usepackage{booktabs}
\usepackage{tabularx}
\usepackage{array}
\usepackage{makecell}

\journal{}

\begin{document}

\begin{frontmatter}

\title{Sagnac-Assisted Enhanced $\phi$-OTDR for Distributed Acoustic Sensing: A Standardized Benchmark and Engineering Evaluation Framework}

\author[aff1]{Weiguang Wang}
\author[aff4,aff2]{Fugen Wu}
\author[aff3]{Hailing Wang}
\author[aff1]{Xuechen Liang}
\author[aff1]{Xiaobin Li}
\author[aff1]{Ru Han}
\author[aff2,aff3]{Tianchang Xie\corref{cor1}}
\ead{Wang0422paper@outlook.com}
\cortext[cor1]{Corresponding author.}

\affiliation[aff1]{organization={East China Jiaotong University},
            addressline={ },
            city={Nanchang},
            postcode={330013},
            state={Jiangxi},
            country={China}}

\affiliation[aff2]{organization={School of Materials and Energy, Guangdong University of Technology},
            addressline={ },
            city={Guangzhou},
            postcode={510006},
            state={ },
            country={China}}

\affiliation[aff3]{organization={Jiangxi Tonghui Technology Group Co., Ltd.},
            addressline={ },
            city={Nanchang},
            postcode={330000},
            state={Jiangxi},
            country={China}}

\affiliation[aff4]{organization={School of Artificial Intelligence and Big Data, Guangzhou Vocational University of Science and Technology},
            addressline={ },
            city={Guangzhou},
            postcode={510550},
            state={ },
            country={China}}

\begin{abstract}
Phase-sensitive optical time-domain reflectometry ($\phi$-OTDR) is widely used in large-scale distributed acoustic sensing (DAS) because it provides distributed spatiotemporal monitoring over long sensing distances. Its field performance can still deteriorate because of polarization-induced fading (PIF), local signal degradation, and strong environmental interference. This study develops a Sagnac-assisted enhanced $\phi$-OTDR sensing architecture and a standardized benchmark framework for engineering-oriented DAS event recognition. The Sagnac interferometer provides a continuous phase response that supplements fading-prone observations in the $\phi$-OTDR channel, and heterogeneous signal alignment is achieved using a cross-correlation procedure implemented on an FPGA platform. The benchmark protocol compares conventional feature-engineering methods, probabilistic shallow classifiers, single-branch deep models, and dual-branch fusion models under consistent data partitioning, preprocessing, and metric definitions. Experiments on a 10-km sensing fiber with six representative acoustic event classes show that the dual-branch fusion model provides the most favorable trade-off among the evaluated methods, reaching 89.79\% accuracy, 89.83\% macro-F1, and a nuisance alarm rate of 5.00\% on the balanced test set. The results also show that channel grouping strongly affects dual-branch evaluation, indicating that deployment-oriented conclusions should be based on accuracy, macro-F1, nuisance alarm rate, false negative rate, and latency rather than accuracy alone. This work provides a physically motivated enhancement strategy for $\phi$-OTDR-based DAS and a reproducible benchmark protocol for future fusion-oriented sensing research.

The implementation and scripts for reproducing the DAS event-recognition experiments are publicly available at https://github.com/wawa-abc/das.
\end{abstract}


\end{frontmatter}


\section{Introduction}

Distributed acoustic sensing (DAS) is a fiber-based sensing technology that enables real-time monitoring of weak disturbances along extended infrastructures and has shown broad application potential in security monitoring, pipeline inspection, and civil infrastructure surveillance \cite{shao2025artificial}. Among existing DAS schemes, phase-sensitive optical time-domain reflectometry ($\phi$-OTDR) has become one of the most widely adopted solutions because it provides distributed spatiotemporal mapping over long sensing distances. Previous studies have sought to improve $\phi$-OTDR performance in terms of signal-to-noise ratio (SNR), spatial resolution, and detection stability \cite{zinsou2019recent}. For example, chirped-pulse-based $\phi$-OTDR has been shown to suppress part of the inherent fading while maintaining system sensitivity \cite{fernandez2019distributed}. Despite these advances, conventional single-configuration $\phi$-OTDR systems remain vulnerable to polarization-induced fading (PIF), local signal degradation, and performance instability under strong environmental interference and harsh field conditions \cite{wang2020adaptability}.

To mitigate the limitations of single-configuration sensing, the Sagnac interferometer has attracted attention as a physically complementary structure. A Sagnac interferometer implemented with balanced polarization-maintaining (PM) fibers can provide a near-zero-delay operating point and improved temperature compensation capability \cite{wada2011balanced}, while hybrid interferometric architectures have further demonstrated advantages in polarization-insensitive design and dynamic range extension \cite{huang2007fiber}. These studies suggest that Sagnac interferometry can provide meaningful physical complementarity to $\phi$-OTDR, especially when polarization-related degradation affects the stability of backscattering-based sensing. Existing work has largely emphasized isolated structural improvements or device-level demonstrations, leaving the system-level evaluation of hybrid architectures under realistic DAS event-recognition scenarios insufficiently explored.

DAS research has also advanced rapidly in both hardware implementation and recognition algorithms. Recent hardware studies have promoted compact, integrated, and lower-cost demodulation architectures, reflecting a clear trend toward scalable practical deployment \cite{yang2025phase, wu2020improved, jin2024silicon}. In parallel, artificial intelligence (AI) and machine learning have been increasingly introduced to improve event recognition and system inference capability \cite{tang2025deep, zhu2023seismic, huang2025review}. Although these methods have improved pattern recognition performance, their practical reliability remains strongly dependent on front-end sensing quality, input organization, and evaluation protocol. The combination of advanced optical demodulation and increasingly sophisticated recognition pipelines has expanded the operational capability of DAS systems, while also making comparisons among different methodological approaches less straightforward. A standardized and reproducible benchmark framework is therefore needed. Tomasov et al. \cite{tomasov2025comprehensive} recently released a comprehensive dataset for validating machine-learning models, while benchmarking studies in fiber-optic strain sensing have shown that different sensing mechanisms exhibit inherent performance trade-offs and that no single technical approach is universally optimal across all scenarios \cite{zensor2026assessing}. These observations further underline the need for a rigorous evaluation framework for advanced DAS demodulation systems.

Motivated by these issues, this paper develops a Sagnac-assisted $\phi$-OTDR sensing architecture and establishes a standardized benchmark evaluation framework for fusion-oriented DAS event recognition. Rather than emphasizing isolated structural optimization or downstream classification accuracy alone, this work connects physical sensing enhancement with engineering-oriented system evaluation. The proposed architecture introduces the continuous phase response of the Sagnac interferometer as an auxiliary source to complement fading-prone observations in the $\phi$-OTDR channel. The evaluation framework shifts the focus from laboratory-style single-metric comparison to a unified benchmark protocol that compares conventional feature-engineering methods, probabilistic shallow classifiers, single-branch deep models, and dual-branch fusion models under consistent data partitioning, preprocessing, and metric definitions. In addition to classification accuracy, the benchmark emphasizes nuisance alarm rate (NAR), false negative rate (FNR), and inference latency to better reflect deployment-oriented performance.

\section{Principle of Sagnac-Assisted $\phi$-OTDR and Benchmark Design}

To address the susceptibility of conventional single-configuration $\phi$-OTDR systems to polarization-induced fading, local signal degradation, and reduced robustness in complex background environments, this work develops a Sagnac-assisted enhanced $\phi$-OTDR sensing architecture together with a standardized benchmark-oriented evaluation framework. The purpose of this paper is not to introduce a new task-specific recognition network, but to establish a physically motivated sensing architecture and a reproducible evaluation protocol for comparing different technical routes under consistent conditions.

The contribution of this section is therefore twofold. First, at the physical layer, the proposed hybrid architecture introduces the high-fidelity continuous phase response of the Sagnac interferometer as an auxiliary sensing source to complement the fading-prone regions of the $\phi$-OTDR channel. Second, at the evaluation layer, the benchmark framework shifts the focus from isolated laboratory accuracy to engineering-oriented reliability, with emphasis on nuisance alarm rate (NAR), false negative rate (FNR), robustness under class imbalance, and fairness of comparison across model families. In this sense, the benchmark is designed as a system-level evaluation framework rather than as a validation vehicle for one specific classifier.

\subsection{Main-Support Architecture and Hybrid Optical Front-End}

\begin{figure}[htbp]
\centering
\includegraphics[width=1.0\textwidth]{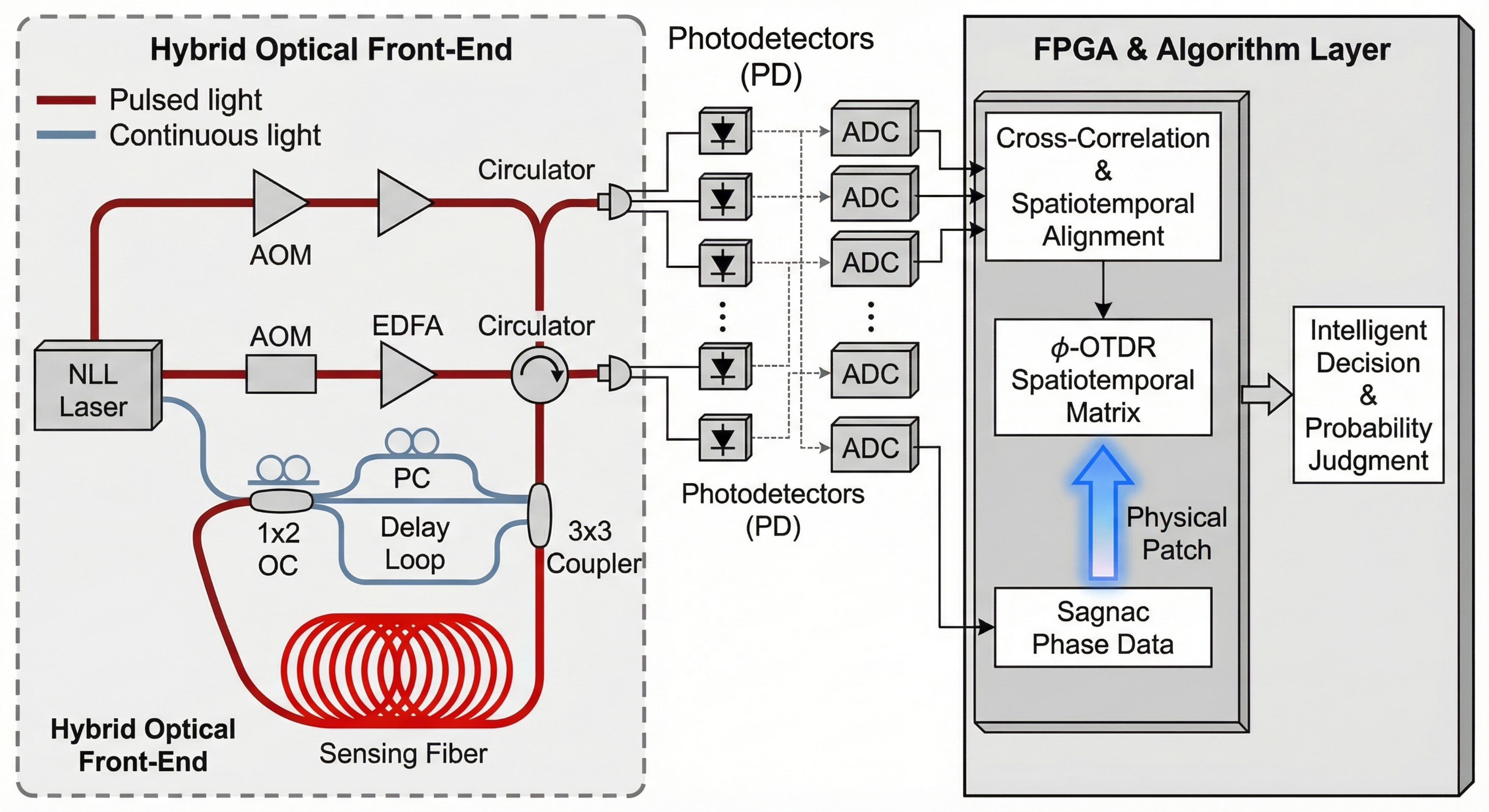}
\caption{Overall architecture of the Sagnac-assisted enhanced $\phi$-OTDR distributed acoustic sensing system, highlighting the hybrid optical front-end and the digital-domain physical enhancement mechanism.}
\label{fig:1}
\end{figure}

The overall architecture of the proposed Sagnac-assisted enhanced $\phi$-OTDR demodulator is illustrated in Fig.~\ref{fig:1}. It consists of a hybrid optical front-end, multi-source signal acquisition, spatiotemporal synchronization, feature construction, and a benchmark-oriented decision layer for downstream event recognition.

The main design principle is to preserve the dominant role of $\phi$-OTDR in distributed spatiotemporal localization while using the Sagnac interference channel as an auxiliary source for compensating locally degraded observations. In other words, $\phi$-OTDR remains responsible for spatially resolved event mapping, whereas the Sagnac channel provides a high-fidelity continuous phase response that is less vulnerable to some local fading effects. This physically complementary arrangement is intended to mitigate the information loss caused by polarization-induced fading (PIF) and strong environmental interference.

From the perspective of the optical sensing mechanism, previous studies have shown that ring Mach--Zehnder or Sagnac interferometric structures can jointly exploit multiple phase responses to estimate both disturbance position and vibration characteristics \cite{sun2008distributed}. However, pure interferometric sensing architectures often have limited long-range distributed localization capability in practical deployments. By contrast, $\phi$-OTDR provides natural distributed sensing through coherent pulse injection and backscattered Rayleigh response acquisition. Its spatial position $z$ can be expressed as
\begin{equation}
z=\frac{c\tau}{2n_{eff}}
\end{equation}
where $c$ is the speed of light in vacuum, $n_{eff}$ is the effective refractive index of the sensing fiber, and $\tau$ is the pulse round-trip time. This physical mapping relationship constitutes the core advantage of $\phi$-OTDR for distributed localization.

Nevertheless, when the Rayleigh backscattering response is degraded by polarization fading or strong local interference, the signal-to-noise ratio at certain spatial positions can drop sharply. Under such conditions, relying solely on downstream pattern recognition to recover information from severely degraded observations may result in unstable decisions, including both missed detections and nuisance alarms. A more reliable solution is therefore to improve the observability of the sensing system at the physical layer. Existing studies have indicated that the joint use of intensity and phase information in coherent DAS systems can compensate for the blind spots of single-dimensional detection and improve recognition robustness \cite{wu2021pattern}.

Motivated by this observation, the present system treats the continuously demodulated Sagnac phase as a physically meaningful auxiliary feature and aligns it with the spatiotemporal response of the $\phi$-OTDR channel. Through cross-correlation-based temporal anchoring and subsequent feature-level organization, the architecture preserves the spatial resolution advantage of $\phi$-OTDR while strengthening the detectability of weak and low-frequency disturbances. In this paper, this mechanism is interpreted as a physical enhancement strategy for sensing robustness rather than as an independent machine-learning contribution.

It should also be clarified that the physical architecture and the benchmark data format are not identical objects. The sensing principle is established on the basis of a Sagnac-assisted $\phi$-OTDR hybrid design, whereas the currently available benchmark dataset is stored as a unified 12-channel matrix rather than as explicitly separated Sagnac and $\phi$-OTDR files. Therefore, the experimental benchmark in the following sections evaluates fusion-oriented recognition routes through controlled channel grouping under the available data format. This setting should be interpreted as a practical and reproducible surrogate evaluation protocol for the proposed hybrid sensing concept, rather than as a strict file-level comparison between two independently stored sensing modalities.

\subsection{Standardized Benchmark Evaluation Framework Design Principles}

Although the hybrid optical front-end provides physical-level enhancement, translating this advantage into engineering reliability requires a rigorous and standardized evaluation framework. Conventional DAS recognition studies are often conducted on relatively balanced laboratory datasets and are commonly judged mainly by overall classification accuracy. However, in real long-distance perimeter monitoring and pipeline surveillance, the distribution of acoustic events is typically highly imbalanced: environmental background disturbances occur much more frequently than true threat events. Under such conditions, accuracy alone is not sufficient to assess whether a sensing system is practically deployable.

For this reason, the benchmark framework proposed in this paper is designed from an engineering perspective rather than from a single-model optimization perspective. Its objective is to support fair comparison among conventional feature-engineering baselines, probability-enhanced shallow classifiers, single-branch deep models, and dual-branch fusion models under the same train--test split, preprocessing procedure, and metric definitions. The benchmark therefore emphasizes reproducibility, comparability, and deployment relevance.

A second design principle is that input organization must itself be treated as a controlled benchmark factor. Because the available dataset is represented as a unified 12-channel input, dual-branch fusion benchmarking is implemented through channel grouping rather than through predefined physically separated sensing files. Consequently, branch construction is not regarded as a neutral implementation detail. Instead, it is explicitly incorporated into the benchmark protocol so that the effect of grouping strategy on fusion performance can be quantitatively evaluated.

A third design principle is that the evaluation target should reflect operational risk. In practical monitoring systems, excessive nuisance alarms reduce usability, whereas missed detections directly affect safety and reliability. Accordingly, the benchmark does not rely on accuracy alone, but evaluates model routes jointly in terms of recognition quality, nuisance alarm rate, false negative rate, and inference latency. This multi-metric setting is intended to reflect the real engineering trade-off faced by industrial DAS deployment.

Finally, the benchmark framework is designed to bridge physical enhancement and algorithmic evaluation without conflating them. The proposed Sagnac-assisted architecture provides the sensing motivation and the physical rationale for heterogeneous information complementarity, whereas the benchmark provides the standardized protocol for testing whether different recognition routes can convert this complementary information into reliable event discrimination. In this way, the benchmark serves as a system-level verification framework for the enhanced sensing architecture, rather than as a dedicated validation framework for any single fusion model.

\section{Dataset and Problem Formulation}

This paper considers six-class DAS event recognition under a unified benchmark setting. Rather than introducing a new sensing architecture or a new fusion network, the objective is to establish a standardized evaluation protocol for comparing conventional feature-engineering methods, probability-enhanced decision models, single-branch convolutional networks, and dual-branch fusion networks under identical data preparation, model execution, and metric definitions.

\subsection{Data Representation and Sensing Background}

In practical optical sensing systems, heterogeneous signals may arise from different physical mechanisms. For example, the output intensity of a Sagnac interferometer can be expressed as
\begin{equation}
I(t)=I_1+I_2+2\sqrt{I_1I_2}\cos[\Delta\phi(t)]
\end{equation}
where $\Delta\phi(t)$ denotes the dynamic phase difference induced by external disturbances. In interferometric sensing, the phase term can be recovered by using an inductive $3\times3$ coupler together with orthogonal phase demodulation \cite{wang2017interferometric}, while polarization control can further improve the fidelity and continuity of the demodulated phase sequence \cite{chen2013elimination}. In addition, real-time acquisition based on field programmable gate array (FPGA) has been widely adopted to support synchronous sampling and processing of heterogeneous sensing streams \cite{martina2019fpga}.

For the main $\phi$-OTDR channel, the spatial position has a natural mapping relationship with the pulse round-trip time:
\begin{equation}
z=\frac{c\tau}{2n_{eff}}
\end{equation}
where $c$ is the speed of light in vacuum, $n_{eff}$ is the effective refractive index, and $\tau$ is the pulse round-trip time. This relationship provides the physical basis for spatiotemporal localization in distributed sensing.

Although the present dataset is organized as a unified 12-channel input rather than explicitly separated Sagnac and $\phi$-OTDR files, the above physical background motivates the benchmark design. In particular, the benchmark retains both single-branch and dual-branch input modes so that different model families can be evaluated in a consistent fusion-oriented setting.

\subsection{Spatiotemporal Alignment}

To ensure comparability across models, the benchmark adopts a standardized alignment procedure before feature extraction and model training. For heterogeneous sensing signals, temporal synchronization is established by maximizing a fragment-level cross-correlation function:
\begin{equation}
R_{sp}(\tau)=\frac{\int_{0}^{T_w}x_s(t)x_p(t+\tau)\,dt}{\sqrt{\int_{0}^{T_w}x_s^2(t)\,dt\int_{0}^{T_w}x_p^2(t)\,dt}}
\end{equation}
where $T_w$ is the sliding window length, $x_s(t)$ denotes the demodulated auxiliary phase sequence, and $x_p(t+\tau)$ denotes the backscattering response extracted from the $\phi$-OTDR spatiotemporal matrix at a given spatial position. By maximizing $R_{sp}(\tau)$, the optimal delay $\tau_{opt}$ can be estimated, thereby establishing a consistent temporal anchor for subsequent analysis \cite{xie2011positioning}.

In this paper, the alignment step is not treated as an independent methodological contribution, but as a standardized front-end operation that guarantees fair comparison among shallow, probabilistic, and deep fusion recognition routes.

\subsection{Task Definition}

The benchmark is formulated as a six-class classification problem. Each sample is represented by a 12-channel signal segment, and the task is to assign the segment to one of the predefined event categories. The same train--test partition and label protocol are used for all benchmarked methods.

To support both single-branch and fusion-based evaluation, the 12-channel input is used in three ways. First, it can be directly fed into a single-branch model. Second, it can be transformed into engineered descriptors for shallow classifiers. Third, it can be divided into two branches for dual-branch fusion models.

This unified formulation allows the benchmark to compare methods of substantially different complexity within the same experimental setting.

\subsection{Evaluation Metrics}

To reflect both recognition quality and deployment relevance, the benchmark reports multiple evaluation metrics rather than accuracy alone. Since the present task is a six-class classification problem, the evaluation protocol distinguishes between multi-class recognition metrics and engineering-oriented alarm metrics.

Let $C=6$ denote the number of classes, and let $M=[m_{ij}] \in \mathbb{R}^{C \times C}$ denote the confusion matrix, where $m_{ij}$ is the number of samples whose true label is class $i$ and predicted label is class $j$. The overall classification accuracy is defined as
\begin{equation}
Accuracy=\frac{\sum_{i=1}^{C} m_{ii}}{\sum_{i=1}^{C}\sum_{j=1}^{C} m_{ij}}.
\end{equation}

For each class $i$, the one-versus-rest precision and recall are defined as
\begin{equation}
Precision_i=\frac{m_{ii}}{\sum_{j=1}^{C} m_{ji}},
\end{equation}
\begin{equation}
Recall_i=\frac{m_{ii}}{\sum_{j=1}^{C} m_{ij}}.
\end{equation}
The class-wise F1-score is then given by
\begin{equation}
F1_i=\frac{2 \cdot Precision_i \cdot Recall_i}{Precision_i + Recall_i}.
\end{equation}

To reduce the influence of class-frequency variation, this paper reports the macro-averaged F1-score:
\begin{equation}
Macro\text{-}F1=\frac{1}{C}\sum_{i=1}^{C} F1_i.
\end{equation}
Unless otherwise specified, all F1 results reported in this paper correspond to macro-F1.

In addition to multi-class recognition metrics, this paper further evaluates engineering-oriented alarm behavior. Since nuisance alarms and missed detections are operational concepts rather than standard multi-class metrics, they are defined through a background-versus-threat aggregation. Specifically, the \texttt{background} class is treated as the non-threat category, whereas all disturbance event classes are jointly treated as threat categories. Let $TP_e$, $TN_e$, $FP_e$, and $FN_e$ denote the event-level true positives, true negatives, false positives, and false negatives under this aggregation. Then the nuisance alarm rate (NAR) and false negative rate (FNR) are defined as
\begin{equation}
NAR=\frac{FP_e}{FP_e+TN_e},
\end{equation}
\begin{equation}
FNR=\frac{FN_e}{FN_e+TP_e}.
\end{equation}

Under this definition, NAR measures the proportion of background samples incorrectly reported as threat events, while FNR measures the proportion of true threat events missed by the system. These two metrics are particularly important for DAS deployment, because excessive nuisance alarms reduce operational efficiency, whereas missed detections directly compromise system reliability.

Finally, latency is measured as the average single-sample inference time. Compared with evaluations based solely on classification accuracy, this multi-metric setting is more suitable for practical DAS deployment, because it explicitly characterizes the trade-off among recognition performance, false alarm suppression, missed-event control, and computational efficiency.

\section{Benchmark Pipeline}

To ensure fair comparison across different technical routes, a unified benchmark pipeline was established for all experiments in this work. The purpose of this pipeline is not to favor one particular classifier, but to guarantee that conventional feature-engineering methods, probability-enhanced shallow models, single-branch deep models, and dual-branch fusion models are evaluated under the same data preparation, preprocessing, and metric definitions. The complete pipeline consists of data standardization, feature construction, model execution, and engineering-oriented output reporting.

\subsection{Data Preprocessing}

All raw samples were first organized under a unified input format in which each event segment is represented by a 12-channel matrix. To reduce inter-sample inconsistency and suppress irrelevant fluctuations, the same preprocessing procedure was applied to all benchmarked routes. Specifically, the pipeline includes baseline detrending, channel-wise normalization, optional downsampling for computational stabilization, and denoising-oriented signal conditioning before subsequent feature extraction or model input construction.

The preprocessing step serves two purposes. First, it reduces the influence of low-frequency drift and channel amplitude scale variation, thereby improving comparability across samples. Second, it ensures that performance differences observed in the benchmark are primarily caused by modeling routes rather than by inconsistent front-end processing. The preprocessing configuration was fixed for all experiments and was not tuned separately for different model families.

\begin{figure}[htbp]
\centering
\includegraphics[width=1.0\textwidth]{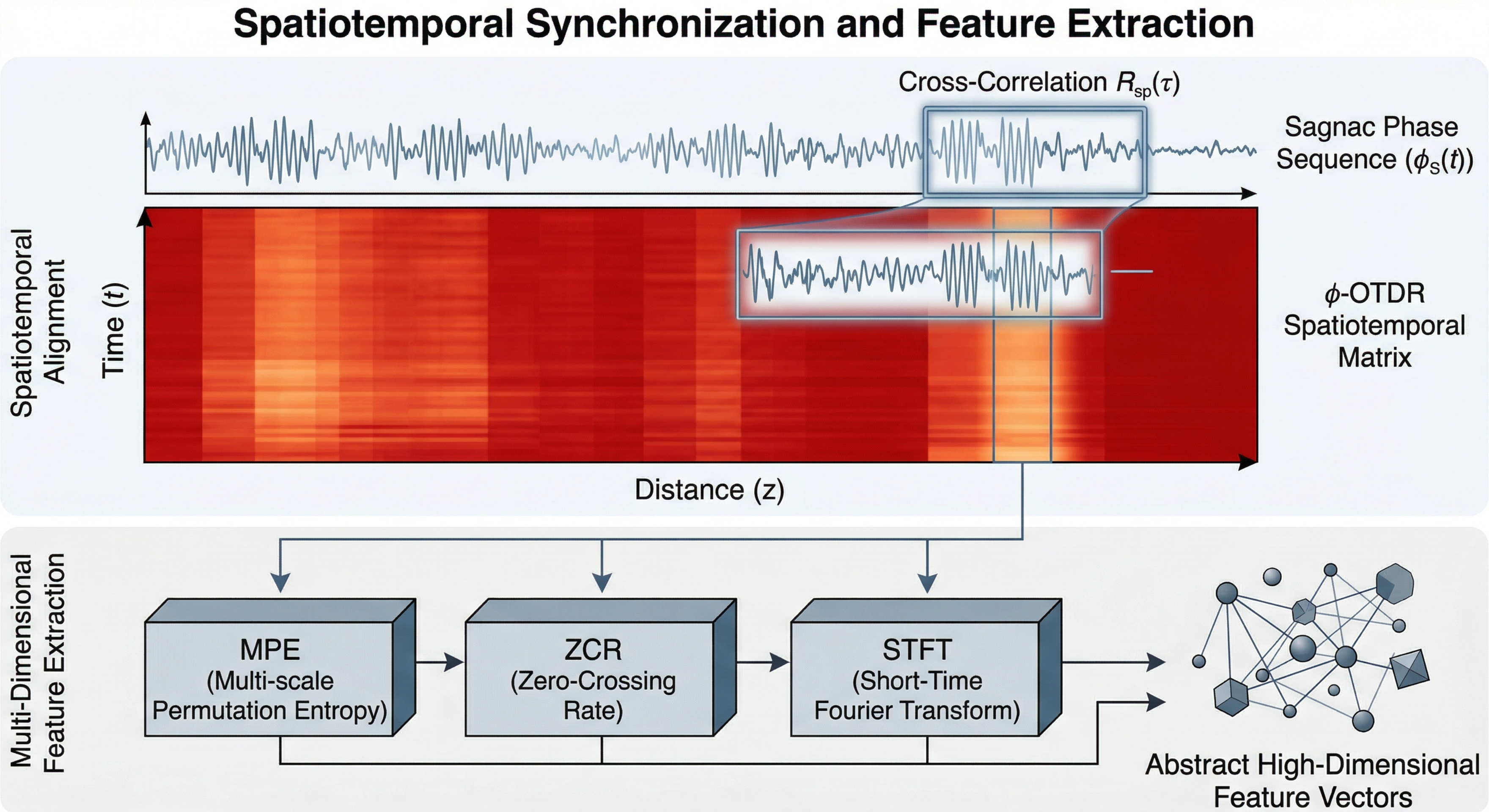}
\caption{Spatiotemporal synchronization and multi-dimensional feature extraction pipeline for the benchmarked DAS event-recognition framework.}
\label{fig:feature_pipeline}
\end{figure}

\subsection{Feature Construction}

To support heterogeneous benchmark routes, the pipeline provides both handcrafted feature construction and direct signal input organization.

For shallow benchmark routes, multiple types of descriptors were constructed from the preprocessed signals. These include time-domain statistical characteristics, short-time Fourier transform (STFT)-based frequency-band energy descriptors, multi-scale permutation entropy (MPE), and zero-crossing rate (ZCR). Time-domain descriptors capture basic amplitude and fluctuation statistics, STFT-based features characterize the frequency distribution of disturbance energy, MPE reflects nonlinear temporal complexity, and ZCR provides a lightweight description of oscillatory behavior. In addition, fusion features were constructed by combining complementary descriptors extracted from multiple channels.

For deep benchmark routes, the preprocessed channel sequences were directly organized as model inputs. In the single-branch setting, the full input or one selected branch was fed to a convolutional network. In the dual-branch setting, the 12-channel input was partitioned into two branches and then processed by a fusion-oriented network. In this way, the benchmark can evaluate both representation learning from raw signals and classification based on engineered descriptors within the same protocol.

\subsection{Benchmarked Model Set}

The benchmark covers four categories of technical routes.

First, conventional shallow single-route methods were included as interpretable reference baselines. These mainly consist of \texttt{STFT + SVM}, \texttt{MPE + ZCR + SVM}, and other lightweight classifier combinations derived from handcrafted descriptors.

Second, probability-enhanced shallow routes were introduced to examine whether posterior-confidence modeling improves the engineering usefulness of shallow classifiers. In this category, fusion-feature-based probabilistic SVM (PSVM) was used as a representative route.

Third, deep single-branch routes were included to evaluate the gain brought by learned task-adaptive representations. These models take one organized signal stream as input and perform end-to-end feature learning through convolutional encoders.

Fourth, deep fusion routes were included to test whether branch-wise complementary information can further improve recognition performance and engineering robustness. In this paper, the fusion CNN should be understood as a representative deep fusion route in the benchmark rather than as the central methodological contribution of the work.

\begin{figure}[htbp]
\centering
\includegraphics[width=1.0\textwidth]{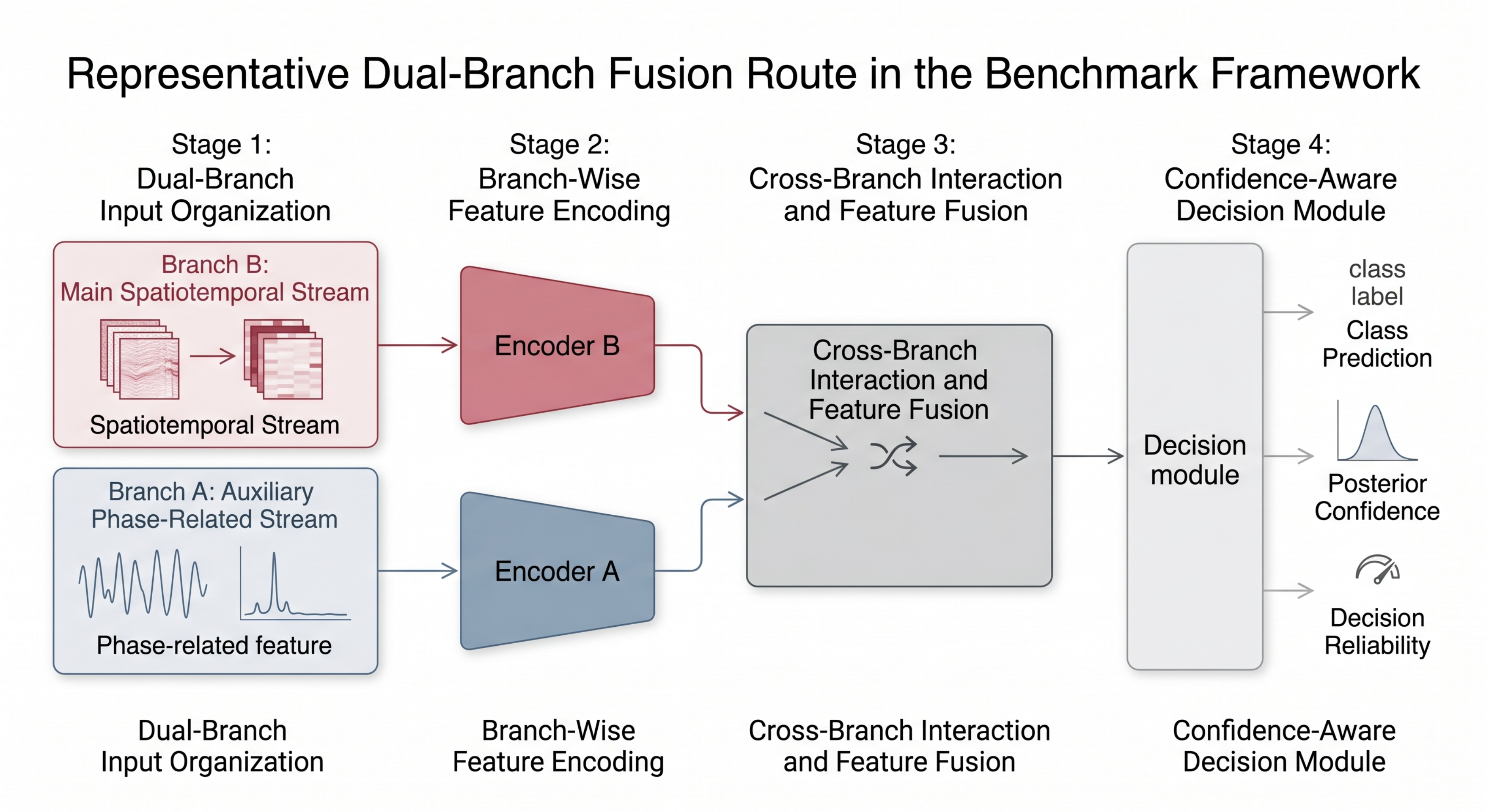}
\caption{Illustration of a representative dual-branch fusion and confidence-aware decision route included in the benchmark comparison. The route consists of dual-branch input organization, branch-wise feature encoding, cross-branch interaction and feature fusion, and confidence-aware decision output.}
\label{fig:representative_fusion_route}
\end{figure}

The benchmarked model set therefore spans the progression from shallow handcrafted descriptors to deep learned representations, and from single-route learning to multi-branch fusion. This design allows the benchmark to reveal the relative performance hierarchy across model complexity levels and information-integration strategies.
Among the benchmarked deep routes, dual-branch fusion provides a representative way to organize complementary information from multiple input streams. Figure~\ref{fig:representative_fusion_route} illustrates a representative fusion route included in the benchmark, where branch-wise encoded features are further integrated through cross-branch interaction before confidence-aware decision output.

It should be emphasized that this figure is used only to illustrate one representative deep fusion route in the benchmark framework rather than to introduce the central methodological contribution of this paper. The purpose of the benchmark is to compare such routes with shallow and single-branch alternatives under the same evaluation protocol.

\subsection{Training and Hyperparameter Configuration}

To avoid unfair comparison caused by inconsistent optimization settings, all benchmarked routes followed a unified training--evaluation protocol. The same train--test split was used for all methods. For conventional shallow models, hyperparameters were selected by standardized search within predefined ranges. For deep models, the training budget, stopping criteria, and evaluation procedure were kept consistent across candidate architectures within the same experimental setting.

Classical models were optimized through conventional parameter search strategies, while deep models were trained under the same data split and monitored by validation-based performance tracking. The objective of this design is not to maximize the absolute performance of one route at all cost, but to maintain fairness and reproducibility in the benchmark comparison.

\subsection{Benchmark Outputs}

For every benchmarked route, the pipeline outputs a standardized result summary including accuracy, macro-F1 score, nuisance alarm rate (NAR), false negative rate (FNR), and single-sample inference latency. In addition, confusion matrices and training convergence curves are generated for representative deep routes. These outputs support both statistical comparison and engineering interpretation.

From the perspective of benchmark construction, this standardized output format is essential. It ensures that all routes are judged not only by recognition accuracy, but also by operationally relevant metrics associated with false alarms, missed detections, and deployment efficiency. As a result, the benchmark can support a more realistic assessment of practical DAS event-recognition systems.

\section{Channel Grouping Optimization}

\subsection{Motivation for Channel Grouping}

Because the currently available benchmark data are stored as a unified 12-channel matrix rather than as explicitly separated Sagnac and $\phi$-OTDR files, dual-branch fusion evaluation cannot rely on a predefined physical branch split. Instead, a controlled channel grouping strategy is required to construct two branches from the same multi-channel input. In this setting, channel grouping is not a trivial implementation detail, but an integral component of the benchmark.

The underlying reason is that the effectiveness of a dual-branch fusion model depends not only on the classifier itself, but also on the degree of complementarity preserved between the two input branches. If two branches contain highly redundant information, the benefit of fusion may be limited. By contrast, if the partition retains stronger heterogeneity and complementarity, the downstream fusion route may exploit richer cross-branch information. Therefore, a fair benchmark must explicitly evaluate the influence of grouping strategy.

\subsection{Candidate Grouping Construction}

To analyze this factor systematically, the 12 input channels were partitioned into two complementary branches of equal size, denoted as \texttt{branch\_a} and \texttt{branch\_b}. Multiple candidate grouping schemes were generated and evaluated under the same protocol. These candidates included default contiguous partitions, interleaved partitions, and additional alternative combinations designed to test whether non-intuitive channel assignments preserve stronger complementarity.

For each candidate grouping, the same preprocessing procedure, model initialization protocol, and evaluation dataset were used. This ensures that differences in performance can be attributed primarily to branch organization rather than to unrelated training conditions. In the present benchmark, 24 candidate groupings were considered in the search stage, and representative results are reported in the experimental section.

\subsection{Search and Ranking Strategy}

The purpose of the grouping search is to identify branch organizations that are most suitable for dual-branch fusion benchmarking. In the search stage, each candidate grouping was first evaluated using the same lightweight screening protocol. The primary ranking score was defined by validation accuracy:
\begin{equation}
Score = Accuracy
\end{equation}
while macro-F1, NAR, and FNR were recorded simultaneously as auxiliary diagnostic criteria.

This design was adopted for two reasons. First, the candidate search stage should remain computationally manageable because multiple grouping schemes must be screened before full benchmark comparison. Second, the final benchmark conclusions are not drawn from search-stage accuracy alone, since the top-ranked groupings are further analyzed together with macro-F1 and engineering metrics in the full evaluation stage. Therefore, the search score provides an efficient first-order ranking, whereas the final benchmark interpretation remains multi-metric.

\subsection{Selection Protocol}

After candidate ranking, the highest-performing grouping schemes were retained for full dual-branch evaluation. The final selection was not based solely on whether one candidate achieved the largest screening accuracy, but also on whether its behavior remained stable under the full benchmark protocol. In this way, grouping optimization was incorporated as a controlled and reproducible step of the benchmark rather than as an arbitrary manual choice.

This protocol improves the fairness of fusion-model comparison. Without explicit grouping optimization, the reported performance of a fusion route could be either underestimated or overestimated depending on the initial branch construction. By contrast, the present procedure allows the benchmark to distinguish between performance gains caused by the fusion model itself and those caused by more suitable input organization.

\subsection{Role in the Benchmark Framework}

The inclusion of channel grouping optimization is one of the key system-level contributions of this work. It reflects the fact that, under the available data format, fusion-oriented DAS benchmarking requires not only classifier comparison but also structured input modeling. In other words, branch construction is part of the benchmark definition.

From a broader perspective, this step also strengthens the engineering relevance of the study. In practical multi-channel sensing systems, the way in which sensing streams are organized and supplied to downstream models can substantially influence final decision quality. Treating grouping as an explicit benchmark factor therefore makes the evaluation framework more rigorous, more reproducible, and more representative of realistic deployment conditions.

\section{Experimental Results and Discussion}

Following the benchmark pipeline and channel grouping optimization strategy described above, the experimental results are organized to compare different technical routes under the same evaluation protocol.

\subsection{Experimental Setup}

\begin{figure}[htbp]
\centering
\includegraphics[width=1.0\textwidth]{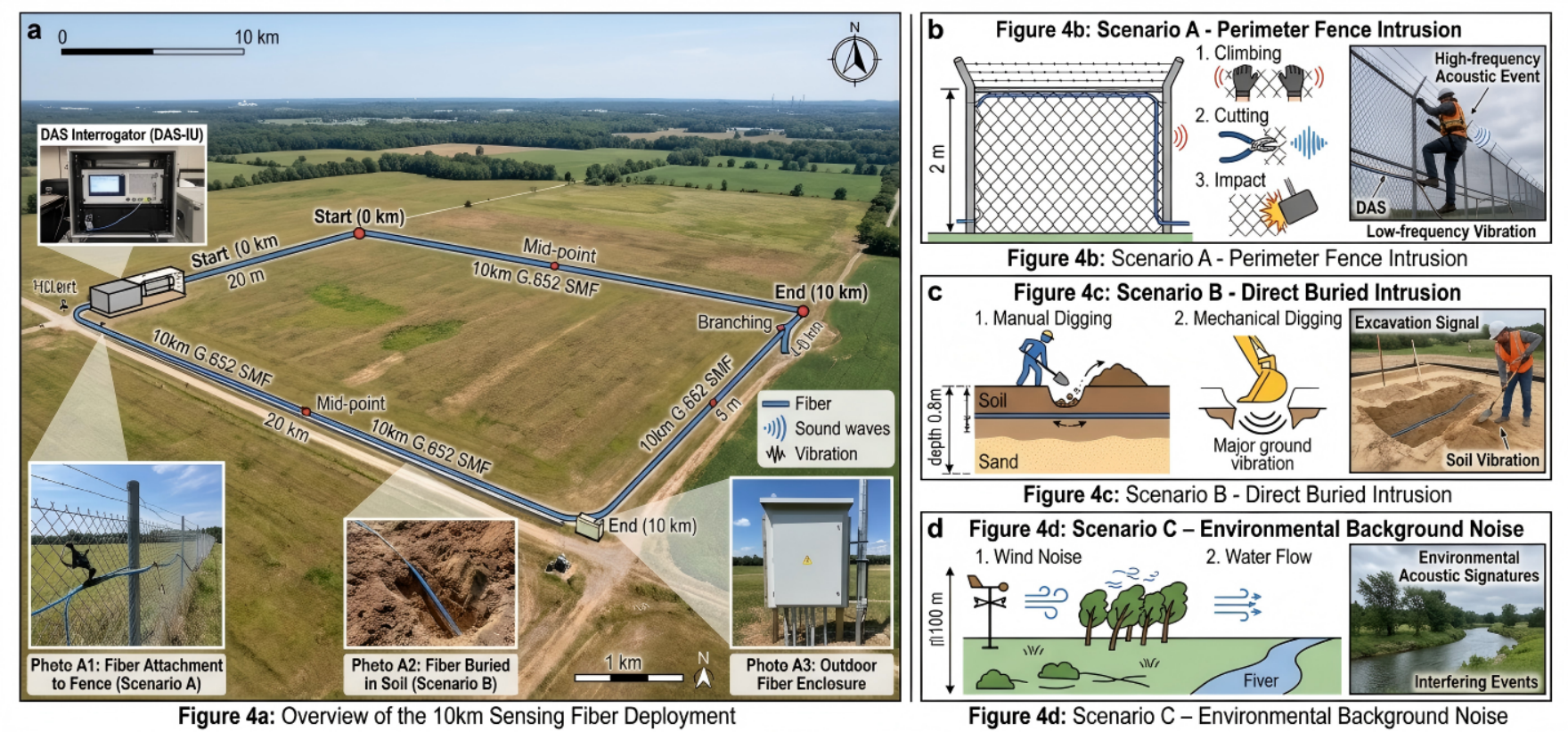}
\caption{Outdoor experimental layout of the 10-km sensing fiber and representative acoustic event simulation scenarios.}
\label{fig:4}
\end{figure}

To validate the proposed benchmark under realistic sensing conditions, experiments were conducted on a 10-km single-mode sensing fiber based on Corning G.652.D. The experimental field was designed to cover representative disturbance scenarios encountered in practical fiber-optic security monitoring. As shown in Fig.~\ref{fig:4}, three typical scenarios were considered. Scenario A corresponds to perimeter intrusion, where the sensing fiber was deployed along a fence structure to record climbing, knocking, and cutting-related disturbances, which typically exhibit transient and high-frequency characteristics \cite{liu2015high}. Scenario B corresponds to pipeline excavation, where the sensing fiber was buried in sandy soil to record manual and mechanical excavation events, whose low-frequency components are prone to attenuation in the underground environment \cite{tejedor2017machine}. Scenario C corresponds to environmental interference, including continuous background disturbances such as wind, water flow, and passing vehicles. Together, these scenarios cover representative event patterns in fiber-optic security systems \cite{bao2012recent}.

To support standardized benchmarking, the dataset was organized into both a balanced dataset and an imbalanced long-tail dataset. The balanced dataset was used for the main benchmark comparison across different technical routes, while the long-tail dataset was used to analyze robustness under severe class imbalance. To avoid data leakage, all samples were split according to independent physical event segments rather than overlapping time windows. In the balanced setting, the full dataset contains 15,419 samples and is divided into training and test sets with an 80:20 ratio. The detailed class distribution is listed in Table~\ref{tab:dataset_distribution}. The data construction procedure, preprocessing protocol, and feature mapping process were fixed throughout the benchmark in order to ensure reproducibility.

\begin{table}[htbp]
\centering
\caption{Data distribution of the balanced benchmark dataset.}
\label{tab:dataset_distribution}
\footnotesize
\renewcommand{\arraystretch}{1.25}
\setlength{\tabcolsep}{3pt}
\begin{tabularx}{0.96\linewidth}{
@{}
>{\raggedright\arraybackslash}m{0.26\linewidth}
>{\raggedright\arraybackslash}m{0.20\linewidth}
>{\centering\arraybackslash}m{0.14\linewidth}
>{\centering\arraybackslash}m{0.14\linewidth}
>{\centering\arraybackslash}m{0.14\linewidth}
@{}
}
\toprule
\makecell[l]{\textbf{Acoustic}\\\textbf{Event}} &
\makecell[l]{\textbf{Class}\\\textbf{Label}} &
\makecell[c]{\textbf{Train Set}\\\textbf{(80\%)}} &
\makecell[c]{\textbf{Test Set}\\\textbf{(20\%)}} &
\makecell[c]{\textbf{Total}\\\textbf{Samples}} \\
\midrule
Background Noise  & 01\_background & 2,357 & 589 & 2,946 \\
Manual Digging    & 02\_dig        & 2,010 & 502 & 2,512 \\
Knocking/Impact   & 03\_knock      & 2,024 & 506 & 2,530 \\
Water Flow        & 04\_water      & 1,802 & 451 & 2,253 \\
Fence Shaking     & 05\_shake      & 2,182 & 546 & 2,728 \\
Walking/Footsteps & 06\_walk       & 1,960 & 490 & 2,450 \\
\midrule
\textbf{Total} & \textbf{-} & \textbf{12,335} & \textbf{3,084} & \textbf{15,419} \\
\bottomrule
\end{tabularx}
\end{table}

Because the available data were stored as a unified 12-channel matrix rather than physically separated primary and auxiliary sensing files, the dual-branch benchmark was implemented through channel grouping. In this way, the 12 channels were partitioned into two branches, and the grouping strategy itself was treated as a controlled benchmark factor rather than a hidden implementation detail. This design preserves the fusion-oriented evaluation framework while avoiding artificial assumptions regarding unavailable physical channel separation.

For robustness evaluation, an extreme long-tail setting was further introduced. The imbalance ratio was defined according to long-term field observations in long-range monitoring tasks, where normal environmental disturbances occur far more frequently than destructive threat events \cite{wu2015distributed}. This imbalanced setting was used to examine whether different benchmarked routes remain reliable when false alarms and missed detections are considered jointly.

All benchmarked methods shared the same train--test split, preprocessing procedure, and evaluation protocol. The benchmark metrics included classification accuracy, macro-F1 score, nuisance alarm rate (NAR), false negative rate (FNR), and single-sample inference latency. The reported results therefore reflect differences in model routes rather than inconsistencies in data preparation. The benchmark includes conventional feature-engineering baselines, probability-enhanced shallow models, single-branch CNNs, and dual-branch fusion CNNs.

\subsection{Overall Benchmark Comparison}

The first objective of the benchmark is to compare shallow, probabilistic, and deep fusion routes under the same experimental protocol. Table~\ref{tab:algorithm_comparison} summarizes the overall performance of representative methods. Conventional feature-engineering baselines, including STFT- and nonlinear-feature-based classifiers, show limited performance on the six-class DAS event recognition task. Specifically, \mbox{STFT + SVM} achieved an accuracy of 41.25\% with a macro-F1 of 38.39\%, while \mbox{MPE + ZCR + SVM} improved the accuracy only slightly to 44.37\%. This indicates that shallow handcrafted features alone are insufficient for capturing the complex intra-class variability and inter-class overlap of practical DAS disturbances.

The introduction of probability-enhanced decision making improved the shallow benchmark route. In particular, the fusion-feature-based PSVM model achieved 56.46\% accuracy and 57.40\% macro-F1, while reducing the false negative rate to 3.25\%, suggesting that posterior probability modeling is beneficial for confidence-aware decision making in noisy sensing environments \cite{cherkassky2004practical,hastie1998classification,zhu2025controllable}. However, even with probabilistic enhancement, shallow models still remained clearly inferior to deep learning routes in overall discrimination capability.

Single-branch CNN baselines substantially outperformed conventional shallow methods. The best single-branch CNN reached 86.04\% accuracy and 85.89\% macro-F1, showing that deep feature extraction is much more effective than handcrafted descriptors for DAS disturbance recognition. Nevertheless, single-branch models remain limited in their ability to exploit complementary information across channel groups.

The dual-branch fusion CNN achieved the most favorable overall engineering trade-off. It reached 89.79\% accuracy and 89.83\% macro-F1, while reducing the nuisance alarm rate to 5.00\% and maintaining a zero false negative rate in the balanced setting. Although its inference latency is higher than that of shallow baselines and some single-branch models, the computational overhead remains acceptable for online monitoring. From the benchmark perspective of this paper, this result is important not because it demonstrates a new architecture, but because it confirms that fusion-based deep routes provide a practically meaningful advantage over both shallow baselines and isolated single-branch models.

\begin{table}[htbp]
\centering
\caption{Benchmark performance comparison of representative methods in the DAS event recognition task.}
\label{tab:algorithm_comparison}
\scriptsize
\renewcommand{\arraystretch}{1.35}
\setlength{\tabcolsep}{3.2pt}
\begin{tabularx}{\linewidth}{
@{}
>{\raggedright\arraybackslash}m{0.17\linewidth}
>{\raggedright\arraybackslash}m{0.25\linewidth}
>{\centering\arraybackslash}m{0.105\linewidth}
>{\centering\arraybackslash}m{0.105\linewidth}
>{\centering\arraybackslash}m{0.085\linewidth}
>{\centering\arraybackslash}m{0.085\linewidth}
>{\centering\arraybackslash}m{0.105\linewidth}
@{}
}
\toprule
\makecell[l]{\textbf{Algorithm}\\\textbf{Architecture}} &
\makecell[l]{\textbf{Feature/Signal}\\\textbf{Route}} &
\makecell[c]{\textbf{Accuracy}\\\textbf{(\%)}} &
\makecell[c]{\textbf{Macro-F1}\\\textbf{(\%)}} &
\makecell[c]{\textbf{NAR}\\\textbf{(\%)}} &
\makecell[c]{\textbf{FNR}\\\textbf{(\%)}} &
\makecell[c]{\textbf{Latency}\\\textbf{(ms)}} \\
\midrule
STFT + SVM &
Manual time--frequency features &
41.25 & 38.39 & 36.25 & 22.00 & 0.0122 \\

MPE + ZCR + SVM &
Nonlinear handcrafted features &
44.37 & 43.35 & 47.50 & 10.75 & 0.0127 \\

Fusion Features + PSVM &
Probability-enhanced shallow fusion &
56.46 & 57.40 & 32.50 & 3.25 & 0.0173 \\

Branch B CNN &
Deep single-branch route &
86.04 & 85.89 & 12.50 & 0.00 & 3.3474 \\

Fusion CNN &
Deep fusion route &
89.79 & 89.83 & 5.00 & 0.00 & 12.7901 \\
\bottomrule
\end{tabularx}
\end{table}

Table~\ref{tab:algorithm_comparison} shows a clear hierarchy in benchmark performance. Conventional handcrafted-feature routes remain weak under complex multi-class disturbance conditions. The probabilistic shallow fusion route improves both macro-F1 and missed-event behavior relative to conventional shallow baselines, but still cannot match the discrimination capability of deep learning routes. The strongest performance is achieved by the deep fusion route, indicating that benchmark conclusions should be based on the joint evaluation of accuracy, macro-F1, NAR, FNR, and latency rather than on one metric alone.

\subsection{Layered Comparison Across Technical Routes}

To better illustrate the performance hierarchy revealed by the benchmark, Table~\ref{tab:route_comparison} organizes the representative methods into four technical routes: shallow single-route methods, probabilistic shallow fusion, deep single-branch learning, and deep fusion.

\begin{table}[htbp]
\centering
\caption{Layered comparison across representative technical routes in the DAS benchmark.}
\label{tab:route_comparison}
\footnotesize
\renewcommand{\arraystretch}{1.3}
\setlength{\tabcolsep}{4pt}
\begin{tabularx}{\linewidth}{
@{}
>{\raggedright\arraybackslash}m{0.25\linewidth}
>{\raggedright\arraybackslash}m{0.24\linewidth}
>{\centering\arraybackslash}m{0.115\linewidth}
>{\centering\arraybackslash}m{0.115\linewidth}
>{\centering\arraybackslash}m{0.095\linewidth}
>{\centering\arraybackslash}m{0.095\linewidth}
@{}
}
\toprule
\makecell[l]{\textbf{Technical}\\\textbf{Route}} &
\makecell[l]{\textbf{Representative}\\\textbf{Method}} &
\makecell[c]{\textbf{Accuracy}\\\textbf{(\%)}} &
\makecell[c]{\textbf{Macro-F1}\\\textbf{(\%)}} &
\makecell[c]{\textbf{NAR}\\\textbf{(\%)}} &
\makecell[c]{\textbf{FNR}\\\textbf{(\%)}} \\
\midrule
Shallow single-route &
MPE + ZCR + SVM &
44.37 & 43.35 & 47.50 & 10.75 \\

Probabilistic shallow fusion &
Fusion Features + PSVM &
56.46 & 57.40 & 32.50 & 3.25 \\

Deep single-branch &
Branch B CNN &
86.04 & 85.89 & 12.50 & 0.00 \\

Deep fusion &
Fusion CNN &
89.79 & 89.83 & 5.00 & 0.00 \\
\bottomrule
\end{tabularx}
\end{table}

A clear progression can be observed in Table~\ref{tab:route_comparison}. Moving from shallow handcrafted descriptors to learned deep representations produces a major performance improvement, while moving further from single-branch learning to branch-wise fusion provides the best overall operational balance. This layered comparison is important because it shows that the benchmark gain is not caused by one isolated design choice, but by a systematic progression from fixed features to learned representations and then to complementary information integration.

First, shallow single-route methods provide interpretable but weak baselines. Their limited performance indicates that manually designed features cannot fully characterize the diverse temporal, spectral, and nonlinear patterns of complex DAS disturbances. Second, probability-enhanced shallow methods improve the confidence structure of decisions and partially alleviate the limitations of hard classification boundaries, but their overall discrimination capability remains constrained by handcrafted representations. Third, deep single-branch CNNs significantly improve recognition performance by learning task-adaptive representations directly from the input signals. Finally, deep fusion methods combine branch-wise complementary information and offer the best overall balance among classification effectiveness, false alarm control, missed-event suppression, and deployment relevance.

Therefore, the benchmark establishes not only which route performs best, but also how the performance hierarchy evolves from shallow fixed-feature modeling to deep fusion-oriented learning.

\subsection{Effect of Channel Grouping and Fusion Organization}

Because the dual-branch benchmark relies on channel grouping rather than physically separated sensing files, the effect of grouping strategy must be evaluated explicitly. Multiple grouping candidates were tested under the same training and evaluation protocol, and representative results are summarized in Table~\ref{tab:grouping_search}. The results show that branch organization has a substantial influence on dual-branch recognition performance.
\begin{table}[htbp]
\centering
\caption{Representative results of channel grouping search for dual-branch fusion benchmarking.}
\label{tab:grouping_search}
\scriptsize
\renewcommand{\arraystretch}{1.35}
\setlength{\tabcolsep}{3.2pt}
\begin{tabularx}{\linewidth}{
@{}
>{\centering\arraybackslash}p{0.055\linewidth}
>{\centering\arraybackslash}p{0.205\linewidth}
>{\centering\arraybackslash}p{0.205\linewidth}
>{\centering\arraybackslash}p{0.115\linewidth}
>{\centering\arraybackslash}p{0.115\linewidth}
>{\centering\arraybackslash}p{0.095\linewidth}
>{\centering\arraybackslash}p{0.095\linewidth}
@{}
}
\toprule
\textbf{Rank} &
\makecell[c]{\textbf{Branch}\\\textbf{A}} &
\makecell[c]{\textbf{Branch}\\\textbf{B}} &
\makecell[c]{\textbf{Accuracy}\\\textbf{(\%)}} &
\makecell[c]{\textbf{Macro-F1}\\\textbf{(\%)}} &
\makecell[c]{\textbf{NAR}\\\textbf{(\%)}} &
\makecell[c]{\textbf{FNR}\\\textbf{(\%)}} \\
\midrule
1 &
\makecell[c]{[1,7,8,\\9,10,11]} &
\makecell[c]{[0,2,3,\\4,5,6]} &
78.75 & 77.46 & 0.00 & 2.00 \\

2 &
\makecell[c]{[2,3,4,\\5,8,9]} &
\makecell[c]{[0,1,6,\\7,10,11]} &
75.83 & 73.40 & 0.00 & 0.00 \\

3 &
\makecell[c]{[0,3,4,\\5,6,11]} &
\makecell[c]{[1,2,7,\\8,9,10]} &
65.83 & 61.32 & 0.00 & 7.50 \\

4 &
\makecell[c]{[0,1,2,\\6,7,8]} &
\makecell[c]{[3,4,5,\\9,10,11]} &
61.67 & 53.72 & 0.00 & 2.00 \\

5 &
\makecell[c]{[0,1,2,\\3,4,5]} &
\makecell[c]{[6,7,8,\\9,10,11]} &
51.25 & 48.22 & 0.00 & 44.50 \\
\bottomrule
\end{tabularx}
\end{table}

Table~\ref{tab:grouping_search} shows that the highest-performing grouping is not the default contiguous partition of the first six versus the last six channels. Instead, the best candidates are obtained by cross-group channel combinations that preserve stronger complementarity between branches. A particularly important observation is that the default front-half versus back-half split yields the weakest result among the representative candidates, with both lower accuracy and substantially worse false negative behavior.

This finding indicates that the effectiveness of a fusion model depends not only on the classifier itself, but also on the degree of complementary information retained by the input grouping strategy. From the perspective of fair comparison, channel grouping should therefore be treated as a controlled benchmark factor. Otherwise, the performance of fusion models may be overestimated or underestimated due to arbitrary branch construction. This observation further supports the need for a standardized benchmark pipeline for fusion-oriented DAS event recognition.

\subsection{Analysis of Confusion Patterns and Representation Behavior}

\begin{figure}[htbp]
\centering
\includegraphics[width=1.0\textwidth]{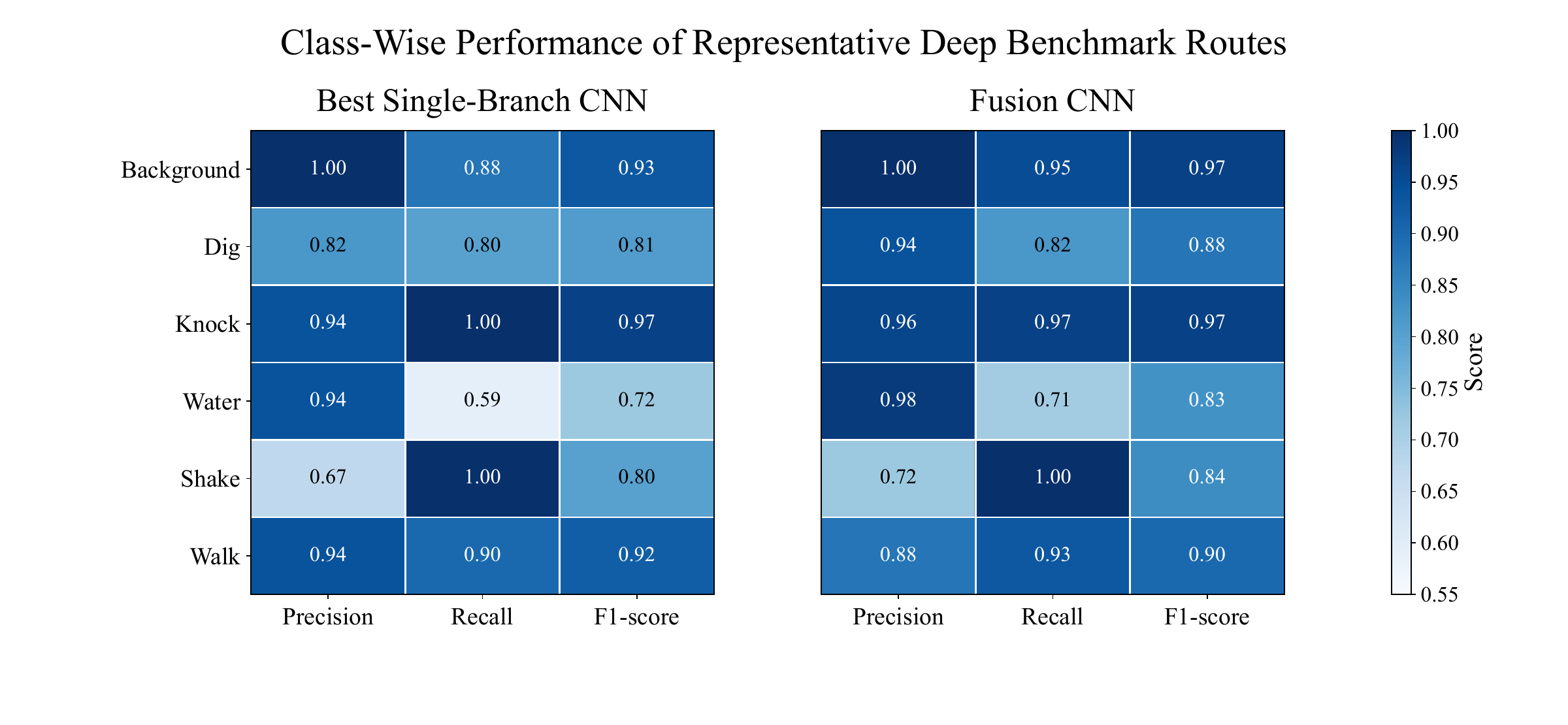}
\caption{Class-wise performance comparison of representative deep benchmark routes in terms of precision, recall, and F1-score.}
\label{fig:5}
\end{figure}

Figure~\ref{fig:5} presents class-wise precision, recall, and F1-score results obtained from representative benchmarked deep routes. Compared with the single-branch models, the fusion route exhibits stronger class-wise recognition behavior for several disturbance categories, indicating that dual-branch interaction helps suppress part of the ambiguity caused by local noise contamination and signal degradation.

More specifically, event categories with relatively stable spectral and temporal signatures, such as \texttt{background} and \texttt{knock}, can already be recognized reliably by single-branch deep models. In contrast, categories more strongly affected by environmental variability, especially \texttt{water} and \texttt{shake}, benefit more from branch fusion. This suggests that the main value of the fusion route lies not in uniformly improving all categories, but in reducing the vulnerability of the system to locally degraded observations.

\begin{figure}[htbp]
\centering
\includegraphics[width=1.0\textwidth]{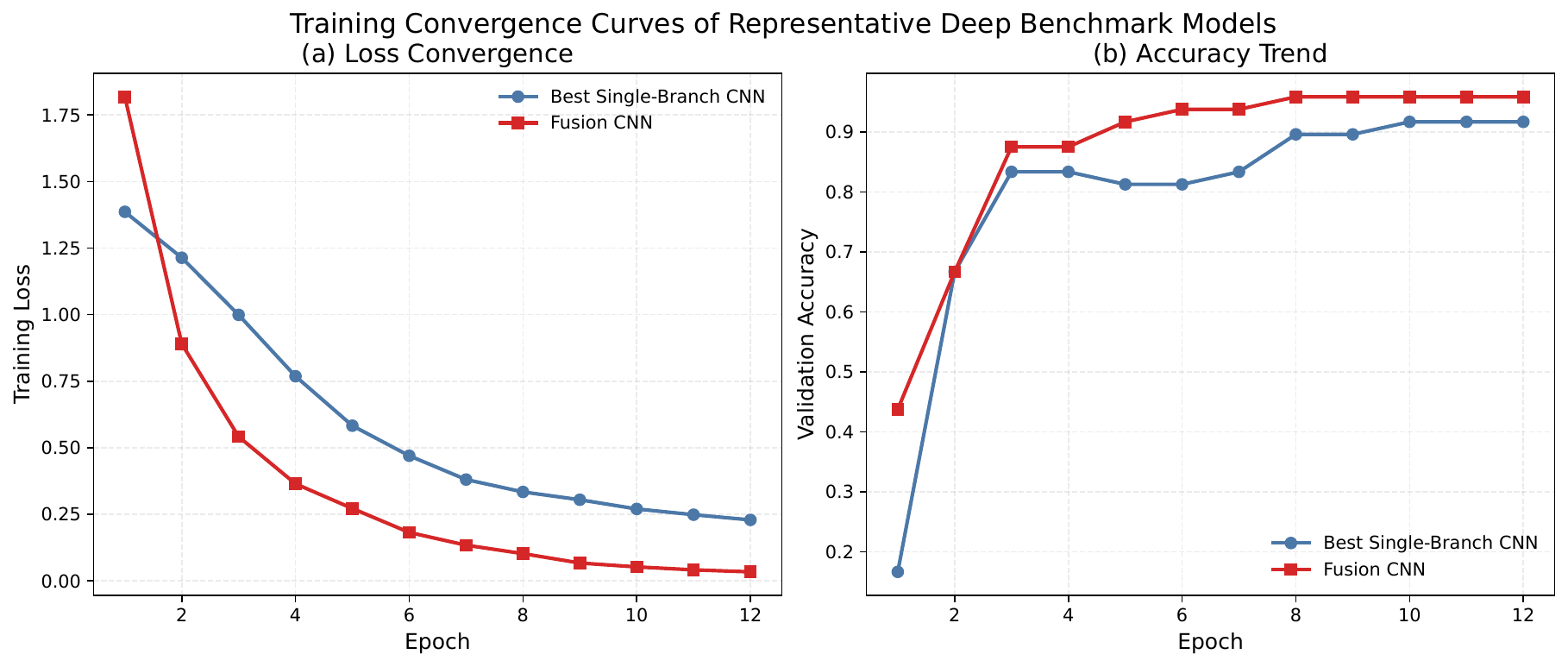}
\caption{Training convergence curves of representative deep benchmark models: (a) training loss and (b) validation accuracy.}
\label{fig:6}
\end{figure}

The convergence behavior of the benchmarked deep models is shown in Fig.~\ref{fig:6}. The training curves indicate that both single-branch and fusion routes are trainable under the same protocol, while the fusion route exhibits a stable optimization process. This supports the validity of the benchmark setting and shows that the reported gains are not caused by unstable optimization artifacts \cite{jiang2019event}.

\subsection{Performance Under Imbalanced Data}

In practical monitoring tasks, destructive events are usually much rarer than background disturbances. Therefore, balanced-dataset accuracy alone is insufficient to assess deployment readiness. To address this issue, the benchmark further evaluates representative technical routes under an imbalanced long-tail setting. In this setting, the class distribution was constructed to reflect the practical tendency that nuisance background signals occur much more frequently than true threat events, while the train--test protocol and evaluation metrics remained consistent with those used in the balanced benchmark.

Because the currently confirmed result files do not yet provide a unified exported benchmark summary covering all four route categories under the same long-tail protocol, the long-tail setting is discussed qualitatively here. The final long-tail numerical summary should be inserted after the standardized pipeline exports the complete results.

Even without a final numerical summary for the long-tail setting, the benchmark design makes the underlying engineering implication clear. Under severe class imbalance, overall accuracy becomes less reliable as a standalone indicator because majority-class dominance can obscure minority-event failure cases. In such scenarios, macro-F1, NAR, and FNR become more informative for judging deployment suitability.

The expected benchmark behavior under this setting is consistent with the balanced-dataset observations. Shallow methods may still retain a certain level of apparent overall accuracy when the majority class dominates the distribution, but their macro-F1 and false negative behavior are generally more vulnerable under severe imbalance. The probabilistic shallow fusion route is more stable than conventional handcrafted baselines, yet its discrimination capability remains limited compared with deep learning routes. By contrast, the deep fusion route is expected to provide the most favorable balance under long-tail conditions because it preserves minority-event recognition while better controlling nuisance alarms from the majority background class.

The long-tail results therefore remain an essential part of the benchmark narrative: models with acceptable balanced-dataset accuracy do not necessarily provide the best operational trade-off under realistic imbalance. Instead, routes that preserve complementary branch information and maintain a more stable balance between NAR and FNR are more suitable for practical security monitoring tasks.

\subsection{Engineering Trade-Offs: NAR, FNR, and Latency}

A major contribution of this benchmark is that it evaluates all methods not only by classification accuracy but also by nuisance alarm rate, false negative rate, and inference latency. This is especially important for DAS deployment, where excessive false alarms reduce operational efficiency and missed detections directly compromise system reliability.

Compared with conventional shallow baselines, deep routes achieve much lower false negative rates. Among the deep models, the fusion route provides the best overall balance between nuisance alarm suppression and missed-event control. Although its single-sample inference latency is higher than that of the single-branch baselines, the additional computational cost remains within an acceptable range for online monitoring. This observation is consistent with prior work showing that practical distributed sensing systems require a balance between recognition performance and real-time execution \cite{qu2019distributed,papp2021real}.

\begin{figure}[htbp]
\centering
\includegraphics[width=1.0\textwidth]{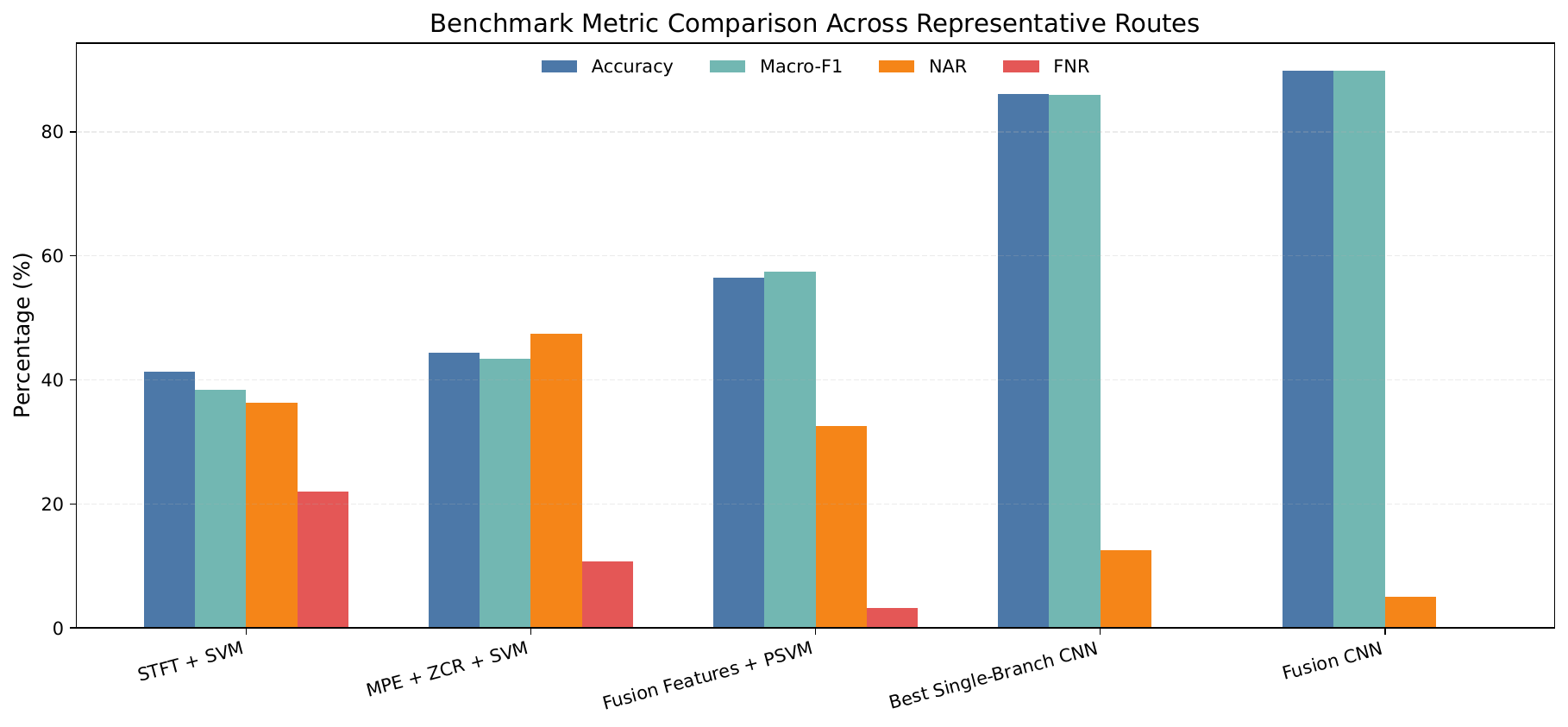}
\caption{Engineering-oriented metric trade-off across representative benchmark routes in terms of accuracy, macro-F1, nuisance alarm rate, and false negative rate.
}
\label{fig:8}
\end{figure}

Figure~\ref{fig:8} further compares engineering-oriented metric trade-offs among representative models. The fusion route achieves a more favorable operating balance than weaker comparison routes, confirming that benchmark conclusions should be supported by multiple deployment-relevant metrics rather than accuracy alone.

Overall, the engineering-oriented evaluation demonstrates that benchmark conclusions should not be drawn from one metric alone. Instead, the deep fusion route is preferred because it offers a better joint balance among recognition quality, nuisance alarm suppression, missed-event control, and latency, whereas shallow baselines remain useful mainly as lightweight reference methods.

\subsection{Discussion}

The benchmark results lead to four main observations. First, conventional feature-engineering methods provide useful interpretability and low-complexity references, but their overall performance is insufficient for complex multi-class DAS event recognition. Second, probability-enhanced shallow methods improve confidence-aware decision making and reduce part of the missed-event risk, yet they still fall short of deep learned representations. Third, deep single-branch CNNs already provide a strong accuracy improvement over shallow routes, confirming the importance of learned task-adaptive representations. Fourth, deep fusion routes deliver the most favorable overall trade-off when classification quality, branch complementarity, nuisance alarm control, missed-event suppression, and latency are considered jointly.

These findings confirm the necessity of a unified benchmark framework for fusion-oriented DAS event recognition. Without a standardized pipeline, conclusions regarding the superiority of one route over another may be confounded by inconsistent preprocessing, unequal data partitioning, arbitrary branch grouping, or incomplete metric reporting. By contrast, the present benchmark provides a reproducible basis for future comparison of conventional, probabilistic, and deep fusion approaches.

At the same time, the present benchmark has a practical boundary that should be acknowledged. The current dual-branch setting is derived from channel grouping over a unified 12-channel matrix rather than from physically independent sensing files. Therefore, the benchmark should be interpreted as a controlled fusion-oriented evaluation framework under the available data format, rather than as a strict comparison between two fully independent sensing modalities. Even with this limitation, the results consistently show that branch organization and fusion strategy are both critical determinants of DAS event recognition performance.

\section{Conclusion}

This paper addressed the limitations of single-configuration $\phi$-OTDR systems in complex engineering environments, where polarization-induced fading (PIF), severe environmental interference, and class-imbalanced event distributions jointly degrade sensing reliability. To overcome these limitations, a Sagnac interference-assisted enhanced $\phi$-OTDR distributed acoustic sensing architecture was developed, and a standardized benchmark framework was established for systematic evaluation under realistic field conditions. Based on the physical design, hardware--software implementation, and multi-route experimental comparison, the main conclusions can be summarized as follows.

First, the Sagnac-assisted physical enhancement mechanism effectively improves the sensing robustness of the conventional $\phi$-OTDR configuration. By preserving the time--space localization capability of $\phi$-OTDR while introducing the high-fidelity continuous phase response of the Sagnac interferometer as an auxiliary sensing source, the proposed architecture alleviates the perception degradation caused by polarization-induced fading. The heterogeneous real-time acquisition and alignment experiments implemented on the FPGA platform confirm that this physical-domain enhancement improves the signal quality of weak and low-frequency disturbance responses, especially under strong background interference.

Second, fusion-oriented representation learning and confidence-aware decision making further improve the engineering reliability of the sensing system. On top of the physical enhancement architecture, this paper constructed a unified benchmark pipeline covering handcrafted feature baselines, probabilistic shallow models, single-branch deep models, and dual-branch fusion CNNs. The experimental results show that shallow handcrafted descriptors alone are insufficient for complex DAS event recognition, whereas deep learned representations provide a clear improvement in classification capability. More importantly, the fusion route yields the most favorable overall trade-off when recognition accuracy, nuisance alarm rate (NAR), false negative rate (FNR), and inference latency are considered jointly. This indicates that the practical value of the proposed system lies not only in improved classification accuracy, but also in better alarm reliability and deployment relevance.

Third, the standardized benchmark protocol established in this work provides a reproducible basis for fair comparison of different technical routes. Under the same train--test split, preprocessing procedure, and evaluation metrics, the proposed benchmark reveals a clear performance hierarchy from shallow single-route models to probabilistic shallow fusion, then to deep single-branch learning, and finally to deep fusion routes. In addition, the results demonstrate that channel grouping and branch organization are not neutral implementation details, but important benchmark factors that substantially influence the effectiveness of dual-branch fusion. This finding is valuable for future research because it clarifies that fair evaluation of fusion-oriented DAS systems requires not only model comparison, but also standardized input organization and engineering-oriented metric reporting.

Overall, the results confirm that the proposed Sagnac-assisted enhanced $\phi$-OTDR framework provides a practically meaningful solution for complex DAS event recognition. Compared with conventional single-route baselines, it offers a more balanced performance in terms of recognition quality, nuisance alarm suppression, missed-event control, and online inference efficiency. Therefore, the contribution of this work is twofold: it improves the sensing robustness of the $\phi$-OTDR architecture at the physical level, and it establishes a unified benchmark framework for evaluating conventional, probabilistic, and deep fusion approaches under realistic monitoring conditions.

Despite these advantages, several issues remain open for future research. First, the current benchmark still relies primarily on supervised learning, whereas real-world monitoring systems often face limited annotation quality and evolving environmental noise. In the future, unsupervised and semi-supervised learning strategies may help reduce dependence on labeled samples and improve the generalization ability of the system for unseen anomalies \cite{zhang2021distributed}. Second, as sensing distance and spatial resolution continue to increase, the resulting spatiotemporal data volume will impose growing pressure on centralized computing architectures. Integrating lightweight models with edge computing platforms is therefore a promising direction for reducing communication overhead and end-to-end inference latency in large-scale fiber sensing networks \cite{wang2022edge}. Third, the present architecture is still constrained by the physical limitations of ultra-long-distance transmission and by the fact that practical infrastructure monitoring often requires joint analysis of vibration, temperature, and strain. Future work can explore online optical amplification, special fiber structures, and unified multi-parameter sensing architectures to further expand the application boundary of enhanced distributed fiber sensing systems \cite{ba2019ultra,dong2020simultaneous,liu2021artificial}.

\bibliographystyle{elsarticle-num}
\bibliography{reference}

@article{shao2025artificial,
  title={Artificial intelligence-driven distributed acoustic sensing technology and engineering application},
  author={Shao, L. and others},
  journal={PhotoniX},
  volume={6},
  number={1},
  year={2025},
  publisher={Springer}
}

@article{zinsou2019recent,
  title={Recent Progress in the Performance Enhancement of Phase-Sensitive {OTDR} Vibration Sensing Systems},
  author={Zinsou, R. and others},
  journal={Sensors},
  volume={19},
  number={7},
  year={2019},
  publisher={MDPI}
}

@article{fernandez2019distributed,
  title={Distributed Acoustic Sensing Using Chirped-Pulse Phase-Sensitive {OTDR} Technology},
  author={Fern{\'a}ndez-Ruiz, M. R. and others},
  journal={Sensors},
  volume={19},
  number={20},
  year={2019},
  publisher={MDPI}
}

@article{wang2020adaptability,
  title={Adaptability and Anti-Noise Capacity Enhancement for {$\phi$-OTDR} With Deep Learning},
  author={Wang, P. and others},
  journal={Journal of Lightwave Technology},
  volume={38},
  number={23},
  year={2020},
  publisher={IEEE}
}

@article{wada2011balanced,
  title={Balanced polarization maintaining fiber {Sagnac} interferometer vibration sensor},
  author={Wada, K. and others},
  journal={Optics Express},
  volume={19},
  number={22},
  year={2011},
  publisher={Optica Publishing Group}
}

@article{huang2007fiber,
  title={Fiber optic in-line distributed sensor for detection and localization of the pipeline leaks},
  author={Huang, S.-C. and others},
  journal={Sensors and Actuators A: Physical},
  volume={135},
  number={2},
  year={2007},
  publisher={Elsevier}
}

@article{zensor2026assessing,
  title={Assessing Reliability of cm-Scale Optical Fiber Strain Sensing in High Gradient Configurations through Benchmarking and Mechanical Modelling},
  author={Zensor and others},
  journal={e-Journal of Nondestructive Testing},
  volume={31},
  number={2},
  year={2026}
}

@article{sun2008distributed,
  title={Distributed fiber-optic vibration sensor using a ring {Mach-Zehnder} interferometer},
  author={Sun, Q. and Liu, D. and Wang, J. and Liu, H.},
  journal={Optics Communications},
  volume={281},
  number={6},
  pages={1538--1544},
  year={2008},
  publisher={Elsevier}
}

@article{wu2021pattern,
  title={Pattern recognition in distributed fiber-optic acoustic sensor using an intensity and phase stacked convolutional neural network with data augmentation},
  author={Wu, H. and Zhou, B. and Zhu, K. and Shang, C. and Tam, H.-Y. and Lu, C.},
  journal={Optics Express},
  volume={29},
  number={3},
  pages={3269},
  year={2021},
  publisher={Optica Publishing Group}
}

@incollection{martina2019fpga,
  title={An {FPGA}-Based Real-Time Acquisition System for a Distributed Acoustic Sensor Based on {$\Phi$-OTDR}},
  author={Martina, F. and others},
  booktitle={Applications in Electronics Pervading Industry, Environment and Society},
  pages={415--420},
  year={2019},
  publisher={Springer}
}

@article{wang2017interferometric,
  title={Interferometric distributed sensing system with phase optical time-domain reflectometry},
  author={Wang, C. and others},
  journal={Photonic Sensors},
  volume={7},
  number={2},
  pages={157--162},
  year={2017},
  publisher={Springer}
}

@article{chen2013elimination,
  title={An Elimination Method of Polarization-Induced Phase Shift and Fading in Dual {Mach--Zehnder} Interferometry Disturbance Sensing System},
  author={Chen, Q. and others},
  journal={Journal of Lightwave Technology},
  volume={31},
  number={19},
  pages={3135--3141},
  year={2013},
  publisher={IEEE}
}

@article{xie2011positioning,
  title={Positioning Error Prediction Theory for Dual {Mach--Zehnder} Interferometric Vibration Sensor},
  author={Xie, S. and others},
  journal={Journal of Lightwave Technology},
  volume={29},
  number={3},
  pages={362--368},
  year={2011},
  publisher={IEEE}
}

@article{cherkassky2004practical,
  title={Practical selection of {SVM} parameters and noise estimation for {SVM} regression},
  author={Cherkassky, V. and Ma, Y.},
  journal={Neural Networks},
  volume={17},
  number={1},
  pages={113--126},
  year={2004},
  publisher={Elsevier}
}

@article{hastie1998classification,
  title={Classification by pairwise coupling},
  author={Hastie, T. and Tibshirani, R.},
  journal={The Annals of Statistics},
  volume={26},
  number={2},
  year={1998},
  publisher={Institute of Mathematical Statistics}
}

@article{zhu2025controllable,
  title={Controllable diffusion framework for imbalanced {Phi OTDR} events classification},
  author={Zhu, B. and others},
  journal={Scientific Reports},
  volume={16},
  number={1},
  year={2025},
  publisher={Nature Publishing Group}
}

@article{liu2015high,
  title={A High-Efficiency Multiple Events Discrimination Method in Optical Fiber Perimeter Security System},
  author={Liu, K. and Tian, M. and Liu, T. and Jiang, J. and Ding, Z. and Chen, Q. and others},
  journal={Journal of Lightwave Technology},
  volume={33},
  number={23},
  pages={4885--4890},
  year={2015},
  publisher={IEEE}
}

@article{tejedor2017machine,
  title={Machine learning methods for pipeline surveillance systems based on distributed acoustic sensing: A review},
  author={Tejedor, J. and Macias-Guarasa, J. and Martins, H. F. and Pastor-Graells, J. and Martin-Lopez, S. and Gonzalez-Herraez, M.},
  journal={Applied Sciences},
  volume={7},
  number={8},
  pages={841},
  year={2017},
  publisher={MDPI}
}

@article{bao2012recent,
  title={Recent progress in distributed fiber optic sensors},
  author={Bao, X. and Chen, L.},
  journal={Sensors},
  volume={12},
  number={7},
  pages={8601--8639},
  year={2012},
  publisher={MDPI}
}

@article{wu2015distributed,
  title={A Distributed Acoustic Sensor for Pipeline Security Monitoring},
  author={Wu, M. and Lu, Y. and Li, J. and Zheng, H. and Peng, W.},
  journal={{IEEE} Photonics Technology Letters},
  volume={27},
  number={18},
  pages={1891--1894},
  year={2015},
  publisher={IEEE}
}

@article{jiang2019event,
  title={An event recognition method for {$\Phi$-OTDR} sensing system based on {CNN}},
  author={Jiang, F. and Tai, H. and Yin, Y. and Liu, C.},
  journal={{IEEE} Sensors Journal},
  volume={20},
  number={3},
  pages={1304--1313},
  year={2019},
  publisher={IEEE}
}

@article{qu2019distributed,
  title={A Distributed Acoustic Sensing System for Perimeter Security Based on {Phase-OTDR} with a Novel Feature Extraction Method},
  author={Qu, Y. and Dong, Y. and Wei, T. and Wu, H. and Zhao, J.},
  journal={Sensors},
  volume={19},
  number={17},
  pages={3773},
  year={2019},
  publisher={MDPI}
}

@article{papp2021real,
  title={Real-time vehicle classification in distributed acoustic sensing using deep learning},
  author={Papp, A. and others},
  journal={Scientific Reports},
  volume={11},
  number={1},
  pages={1--12},
  year={2021},
  publisher={Nature Publishing Group}
}

@article{zhang2021distributed,
  title={Distributed acoustic sensing using semi-supervised learning for anomaly detection},
  author={Zhang, J. and others},
  journal={{IEEE} Internet of Things Journal},
  volume={8},
  number={20},
  pages={15383--15392},
  year={2021},
  publisher={IEEE}
}

@article{wang2022edge,
  title={Edge computing for real-time distributed acoustic sensing data processing},
  author={Wang, X. and others},
  journal={Journal of Lightwave Technology},
  volume={40},
  number={10},
  pages={3290--3298},
  year={2022},
  publisher={IEEE}
}

@article{ba2019ultra,
  title={Ultra-long-distance phase-sensitive optical time-domain reflectometry},
  author={Ba, D. and others},
  journal={Optics Express},
  volume={27},
  number={10},
  pages={14143--14152},
  year={2019},
  publisher={Optica Publishing Group}
}

@article{dong2020simultaneous,
  title={Simultaneous measurement of vibration and temperature based on distributed optical fiber sensor},
  author={Dong, Y. and others},
  journal={{IEEE} Photonics Technology Letters},
  volume={32},
  number={11},
  pages={671--674},
  year={2020},
  publisher={IEEE}
}

@article{liu2021artificial,
  title={Artificial intelligence in distributed optical fiber sensing: A review},
  author={Liu, H. and others},
  journal={Sensors},
  volume={21},
  number={16},
  pages={5394},
  year={2021},
  publisher={MDPI}
}

@article{yang2025phase,
  title={Phase demodulation of hybrid $3\times3$ coupler and Sagnac interferometer for $\phi$-OTDR},
  author={Yang, Binyuan and Wang, Tingyu and Zhang, Jianzhong and Ma, Zhe and He, Xiang and Liu, Lipu and Wang, Yixuan and Zhang, Mingjiang},
  journal={Frontiers in Physics},
  volume={13},
  pages={1609493},
  year={2025},
  publisher={Frontiers}
}

@article{wu2020improved,
  title={The improved denoising algorithm of acoustic sensor based on linear optical fiber Sagnac interferometer},
  author={Wu, Huaming and He, Le and Chen, Hepu and Xiao, Wenbo and Guo, Zhuang and Duan, Junhong and He, Xingdao},
  journal={Optical Fiber Technology},
  volume={60},
  pages={102363},
  year={2020},
  publisher={Elsevier}
}

@article{jin2024silicon,
  title={Silicon photonic integrated interrogator for fiber-optic distributed acoustic sensing},
  author={Jin, Zhicheng and Chen, Jiageng and Chang, Yanming and Liu, Qingwen and He, Zuyuan},
  journal={Photonics Research},
  volume={12},
  number={3},
  pages={465},
  year={2024},
  publisher={Optica Publishing Group}
}

@article{tang2025deep,
  title={Deep learning-based phase demodulation for distributed acoustic sensor},
  author={Tang, Yiming and Liu, Kewei and Liu, Chen and Wu, Haiyong and Wang, Rugang and Chen, Mengmeng and Xu, Fei},
  journal={Scientific Reports},
  volume={15},
  number={1},
  pages={29767},
  year={2025},
  publisher={Nature Publishing Group}
}

@article{zhu2023seismic,
  title={Seismic arrival-time picking on distributed acoustic sensing data using semi-supervised learning},
  author={Zhu, Weiqiang and Biondi, Ettore and Li, Jiaxuan and Yin, Jiuxun and Ross, Zachary E and Zhan, Zhongwen},
  journal={Nature Communications},
  volume={14},
  number={1},
  pages={8192},
  year={2023},
  publisher={Nature Publishing Group}
}

@inproceedings{huang2025review,
  title={A Review of Distributed Acoustic Sensing (DAS) for High-Voltage Power-Cable Fault Monitoring},
  author={Huang, Yuping and Ma, Changwei and Zhang, Jue and Liu, Ruipeng and Chen, Wenjiao and Hu, Shangyu and Peng, Fei and Miao, Qiang},
  booktitle={2025 International Conference on Sensing, Measurement \& Data Analytics in the era of Artificial Intelligence (ICSMD)},
  pages={1--6},
  year={2025},
  organization={IEEE}
}

@article{tomasov2025comprehensive,
  title={Comprehensive Dataset for Event Classification Using Distributed Acoustic Sensing (DAS) Systems},
  author={Tomasov, Adrian and Zaviska, Pavel and Dejdar, Petr and Klicnik, Ondrej and Horvath, Tomas and Munster, Petr},
  journal={Scientific Data},
  volume={12},
  number={1},
  pages={793},
  year={2025},
  publisher={Nature Publishing Group}
}

\end{document}